\documentclass[epj]{webofc}
\usepackage[utf8]{inputenc}
\usepackage[varg]{txfonts}   % Web of Conferences font
\usepackage{booktabs}
\usepackage{xcolor}
\definecolor{darkred}{rgb}{0.4,0.0,0.0}
\definecolor{darkgreen}{rgb}{0.0,0.4,0.0}
\definecolor{darkblue}{rgb}{0.0,0.0,0.4}
\usepackage[bookmarks,linktocpage,colorlinks,
    linkcolor = darkred,
    urlcolor  = darkblue,
    citecolor = darkgreen]{hyperref}
\usepackage{subfigure}
\wocname{EPJ Web of Conferences}
\woctitle{Lattice2017}
\newcommand{\beq}{\begin{eqnarray}}
\newcommand{\eeq}{\end{eqnarray}}

\newcommand{\VG}{\langle G^2 \rangle}
\def\eq#1{Eq.~(\ref{#1})}
\begin{document}
%%%%%%%%%%%%%%%%%%%%%%%%%%%%%%%%%%%%%%%%%%%%%%%%%%%%%%%%%%%%%%%%%%%%%%%%%%%%%
%
\selectlanguage{english}
%----------------------------------------------------------------------------
\title{% indico...
Instanton dominance over $\alpha_s$ at low momenta from lattice QCD simulations at $N_f=0$, $N_f=2+1$ and $N_f=2+1+1$}
%----------------------------------------------------------------------------
\author{% A. , Ph. , F. De Soto, J. Rodr\'{\i}guez-Quintero, S. }
\firstname{Andreas} \lastname{Athenodorou}\inst{1} \and
\firstname{Philippe} \lastname{Boucaud}\inst{2} \and
\firstname{Feliciano} \lastname{de Soto}\inst{3,8}\fnsep\thanks{Speaker, \email{fcsotbor@upo.es}}\and
\firstname{Jos\'e} \lastname{Rodr\'{\i}guez-Quintero}\inst{4,8} \and
\firstname{Savvas} \lastname{Zafeiropoulos}\inst{5,6,7} 
% etc.
}
%----------------------------------------------------------------------------
\institute{%1
Computation-based Science and Technology Research Center, Cyprus Institute, 20 Kavafi Str., Nicosia 2121, Cyprus
\and %2
Laboratoire de Physique Th\'eorique (UMR8627), CNRS,
Univ. Paris-Sud, Universit\'e Paris-Saclay, 91405 Orsay, France
\and %3
Dpto. Sistemas F\'{\i}sicos, Qu\'{\i}micos y Naturales, 
Univ. Pablo de Olavide, 41013 Sevilla; Spain
\and %4
Dpto. Ciencias Integradas, Fac. Ciencias Experimentales; 
Universidad de Huelva, 21071 Huelva; Spain.
\and %5
Thomas Jefferson National Accelerator Facility, Newport News, VA 23606, USA\and%6
Department of Physics, College of William and Mary, Williamsburg, VA 23187-8795, USA\and%7
Institute for Theoretical Physics, Universit\"at Heidelberg, Philosophenweg 12, D-69120 Germany%8
\and %9
CAFPE, Universidad de Granada, E-18071 Granada, Spain
}
%----------------------------------------------------------------------------
\abstract{% the indico one...
We report on an instanton-based analysis of the gluon Green functions in the Landau gauge for low momenta; in particular we use lattice results for $\alpha_s$ in the symmetric momentum subtraction scheme (${\rm MOM}$) for large-volume lattice simulations. We have exploited quenched gauge field configurations, $N_f=0$, with both Wilson and tree-level Symanzik improved actions, and unquenched ones with $N_f=2+1$ and $N_f=2+1+1$ dynamical flavors (domain wall and twisted-mass fermions, respectively).

We show that the dominance of instanton correlations on the low-momenta gluon Green functions can be applied to the determination of phenomenological  parameters of the instanton liquid and, eventually, to a determination of the lattice spacing.

We furthermore apply the Gradient Flow to remove short-distance fluctuations. The Gradient Flow gets rid of the QCD scale, $\Lambda_{\rm QCD}$, and reveals that the instanton prediction extents to large momenta. For those gauge field configurations free of quantum fluctuations, the direct study of topological charge density shows the appearance of large-scale lumps that can be identified as instantons, giving access to a direct study of the instanton density and size distribution that is compatible with those extracted from the analysis of the Green functions.}
%----------------------------------------------------------------------------
\maketitle
%----------------------------------------------------------------------------
% 
% Outline of the proceeding
%
% IR of thermalised green functions Nf=0,2+1,2+1+1 (not details on the lattice configurations, only DWF is new)
% Localization, instanton size distribution (histogram)
% For the cold confs they coincide within a large uncertainty
%
% Only in the paper (not here):
% 	model density evolution
%	pair distribution functions
% 	detailed description of WF
%
% FOCUS	in the evolution with WF that requires for a more detailed scrutiny
%	maybe show only the histogram, not Fig.2 with tau evolution...
%

\section{Introduction}\label{intro}

Instantons and other semi-classical solutions of the QCD Lagrangian are believed to play an essential role in the low energy dynamics of QCD, where the crucial phenomena of confinement and chiral symmetry breaking take place~\cite{RevModPhys.70.323}. Instantons are intimately related to the topological properties of the QCD gauge fields, inherently non-perturbative and one can study their effects via lattice simulations. Phenomenologically, they are very important since among others they can provide a resolution to the U(1) problem, provide a mechanism of chiral symmetry breaking, they can be responsible for rare decays of baryons.

For any gauge field configuration, both the topological charge and the instanton contribution to the lattice gauge fields are hidden by the presence of short-range (UV) fluctuations and most studies reveal the presence of instantons only after the application of a filtering technique. Cooling, different smearing methods and, more recently, Wilson flow are efficient techniques that remove short-range fluctuations in the gauge fields that have been widely exploited for the study of QCD topology~\cite{PhysRevD.89.105005,PhysRevD.92.125014}. 

After UV fluctuations have been removed, different methods have been used to recognize instantons in the resulting gauge fields~\cite{PhysRevD.52.4691,PhysRevD.78.054506,PhysRevD.88.034501} and measure their density and size distribution.
The application of a filtering technique may nevertheless introduce biases in the characteristics of the underlying semiclassical configuration. For example, instanton/anti-instanton pair annihilation would lead to a reduction of the instanton density that would therefore depend on the amount of UV filtering applied. The fact that the instanton size may be modified by the filtering is also known, at least with the use of cooling, in such a way that depending on the gauge action used, instantons shrink or grow with cooling. A third phenomenon, is the disappearance of small instantons of size comparable to the lattice spacing which may introduce uncontrolled effects, affecting the measure of the topological charge. %aqui falta citar a M. Garcia-Perez, Van Baal, ...

A different approach to the determination of the instanton nature of QCD vacuum was presented in \cite{JHEP04Boucaud}, where the IR running of some gluon Green functions was asserted to be related to some properties of the instanton ensemble. Much effort has been devoted by the non-perturbative QCD community to the understanding of the deep IR running of the correlation functions among the fundamental fields of the theory. Quite remarkably, the combination of lattice methods and Dyson-Schwinger equations results has led to the firm conclusion that the gluon propagator acquires a dynamical mass in the deep IR while ghost propagator remains massless. An appealing possibility to describe the large distance (low momenta) correlations among gluon fields is the use of an instanton liquid model with the advantage that it can be applied both before and after removal of UV fluctuations. 

Our goal is to apply both the direct recognition of instantons and the analysis of the running of gluon Green functions, using the comparison between both methods in order to quantify the systematic uncertainties. Overall we expect to obtain a precise image of the instanton description of the QCD vacuum, and explore the dependence of the instanton liquid parameters with the number of dynamical fermions of the theory, by exploiting a large number of quenched ($N_f=0$) and unquenched (with $N_f=2+1$ and $N_f=2+1+1$ dynamical flavors) configurations.

For the quenched data, we considered the large volume simulations employed in \cite{PhysRevD.95.114503,Pepe_lat2017} for a very detailed study of the deep IR running of gluon Green functions, while for the unquenched case we will use two sets of lattice data. The first includes $N_f=2+1$ field configurations produced by the RBC/UKQCD collaboration using domain wall fermions close to the physical pion mass~\cite{PhysRevD.93.074505} and three different values of $\beta$. 590 configurations for $\beta=2.37$ ($a=0.063{\rm fm}$, $V=(2.0{\rm fm})^4$), 330 configurations  for $\beta=2.25$ ($a=0.084{\rm fm}$, $V=(5.3 {\rm fm})^4$) and 350 configurations  for $\beta=2.13$ ($a=0.114{\rm fm}$, $V=(5.5{\rm fm})^4$). We will also use 200 gauge field configurations with $N_f=2+1+1$ dynamical flavors from the ETM collaboration \cite{Baron2010} with a pion mass of around $300$ MeV at $\beta=1.90$ ($a=0.084{\rm fm}$, $V=4.0^3\times 7.9 {\rm fm}^4$). 

%0 & Tree-level Symanzik & 3.8     & $0.285 {\rm fm}$        & $(13.7 {\rm fm})^4$       & 1050  & \cite{ATHENODOROU2016354} \\
%  &                  & 3.9     & $0.243 {\rm fm}$        & $(15.6 {\rm fm})^4$       & 2000  &  \\
%  &                  & 4.2     & $0.141 {\rm fm}$        & $(4.5 {\rm fm})^4$        & 420   &  \\ 
%  &  Wilson          & 5.8     & $0.140 {\rm fm}$        & $(6.72 {\rm fm})^4$       & 960   &  \\
%  &                 & 5.6     & $0.235 {\rm fm}$        & $(11.3 {\rm fm})^4$       & 1920  &  \\
%  &                 & 5.6     & $0.235 {\rm fm}$        & $(12.3 {\rm fm})^4$       & 1790  &  \\
%\hline
%2+1 &  Domain-Wall fermions & 2.37     & $0.0626 {\rm fm}$        & $(2.00 {\rm fm})^4$       & 590  & \cite%{PhysRevD.93.074505}  \\
%  &           & 2.25     & $0.0835 {\rm fm}$        & $(5.34 {\rm fm})^4$       & 330  &   \\
%  &           & 2.13     & $0.1139 {\rm fm}$        & $(5.48 {\rm fm})^4$       & 350  &   \\
%\hline
%2+1+1 & Twisted-mass fermions    & 1.95     & $0.083 {\rm fm}$        & $4.0^3\times 7.9 {\rm fm}^4$       &   200 & \cite{Baron2010}\\

\section{The IR running of gluon Green functions} 
\label{sec:green}

Instantons are thought to describe long distance physics of QCD and therefore they may play a central role in the running of QCD Green functions at low momenta. A simple prediction from an instanton liquid model was proposed some years ago~\cite{PhysRevD.70.114503,JHEP03Boucaud} to detect the instantonic nature of gluon correlation functions at low momenta. This method, that was recently applied to unquenched lattice data~\cite{ATHENODOROU2016354,Athenodorou:2018jwu}, will be depicted below, and presents the advantage of offering a procedure to study the instantonic content of lattice gauge field configurations without filtering short-range fluctuations. 

\subsection{Instanton prediction}

In many phenomenological studies which employ a semi-classics approach one constructs an approximate solution as the superposition of instantons and anti-instantons placed at different positions. This type of ansatz can be found in ~\cite{Shuryak:1987iz,ATHENODOROU2016354}. Due to lack of space we will avoid repeating all the relevant formulae of the instanton liquid analysis and we will refer the reader to ~\citep{ATHENODOROU2016354} for all the details. 
%as
%\beq
%g A_\mu^a (x) = \sum_i 2 R^{\alpha a}_i \bar{\eta}^\alpha_{\mu\nu} \frac{y_i^\nu}{y_i^2} \phi\left(\frac{|y_i|}{\rho_i}\right)
%\label{eq:sumansazt}
%\eeq
%with $y_i=x-x_0^i$ and  $x_0^i$ the instanton center, $R^{\alpha a}_i$ a rotation matrix in color space used to embed the SU(2) instanton into SU(3), $\bar\eta^\alpha_{\mu\nu}$ t'Hooft symbol ($\eta^\alpha_{\mu\nu}$ for anti-instantons) and $\rho$ a length parameter known as instanton radius or instanton size. This ansatz depends on the profile function $\phi(z)$ that controls the dependence of the gauge fields on the distance from the instanton center. Near each instanton center, this profile function is thought to behave as an isolated BPST instanton~\cite{BELAVIN197585}: 
%\beq
%\phi_{BPST}(z)=\frac{1}{1+z^2}
%\eeq
%while for large distances a faster falloff is expected due to instanton superposition or interaction~\cite{PhysRevD.70.114503}.
%
%If the QCD vacuum is described by \eq{eq:sumansazt}, the scalar form factor of the gluon propagator in Landau gauge would be given by~\cite{JHEP03Boucaud}:
%\beq
%G^{(2)}(k^2) = \frac{n}{8g_0^2} \langle \rho^6 I^2(k\rho) \rangle
%\eeq
%where the dependency on the profile function $\phi(z)$ appears through
%\beq
%I(s) = \frac{8\pi^2}{s} \int_0^\infty dz\ z J_2(s z) \phi(z) \ ,
%\eeq
%$n$ is the instanton density, and $\langle\cdots \rangle$ stands for the average over instanton sizes. The scalar form factor for the symmetric three-gluon vertex writes similarly as
%\beq
%G^{(3)}(k^2) &=& \frac{n}{48 k g_0^3} \langle \rho^9 I^3(k\rho) \rangle \ .
%\eeq

The strong coupling constant in the symmetric momentum substraction scheme (MOM), defined from the  two- and three-gluon scalar form factors, and its counterpart in an instanton background are 
\beq
\alpha_{\rm MOM}(k) = \frac{k^6}{4\pi} \frac{\left(G^{(3)}(k^2)\right)^2}{\left(G^{(2)}(k^2)\right)^3}\qquad \mbox{ and  } \qquad \alpha_{\rm MOM}(k) = \frac{1}{18\pi n} \frac{\langle \rho^9 I(k\rho)^3\rangle^2}{\langle \rho^6 I(k\rho)^2\rangle^3}\ .
\label{eq:mom}
\eeq

The most remarkable feature of this prediction is that in the limits of both small and large momenta, this ratio of Green functions becomes independent of the instanton profile~\cite{JHEP04Boucaud,ATHENODOROU2016354}, keeping only a dependence on the instanton density
\beq
\alpha_{\rm MOM}(k) = \frac{c k^4}{18\pi n} \ .
\label{eq:k4}
\eeq
The constant $c$ takes the value $1$ for $k\bar\rho\gg 1$ (with $\bar\rho=\sqrt{\langle\rho^2\rangle}$) and is related to the instanton size distribution for $k\bar\rho\ll 1$. At first order in the mean square width of the size distribution, $\delta\rho^2=\langle(\rho-\bar\rho)^2\rangle$, it can be approximated by $c\approx1+48\frac{\delta\rho^2}{\bar\rho^2}$.

\subsection{Lattice results}

\begin{figure}[h!]
\begin{center}
\begin{tabular}{c c c}
\includegraphics[width=0.32\textwidth]{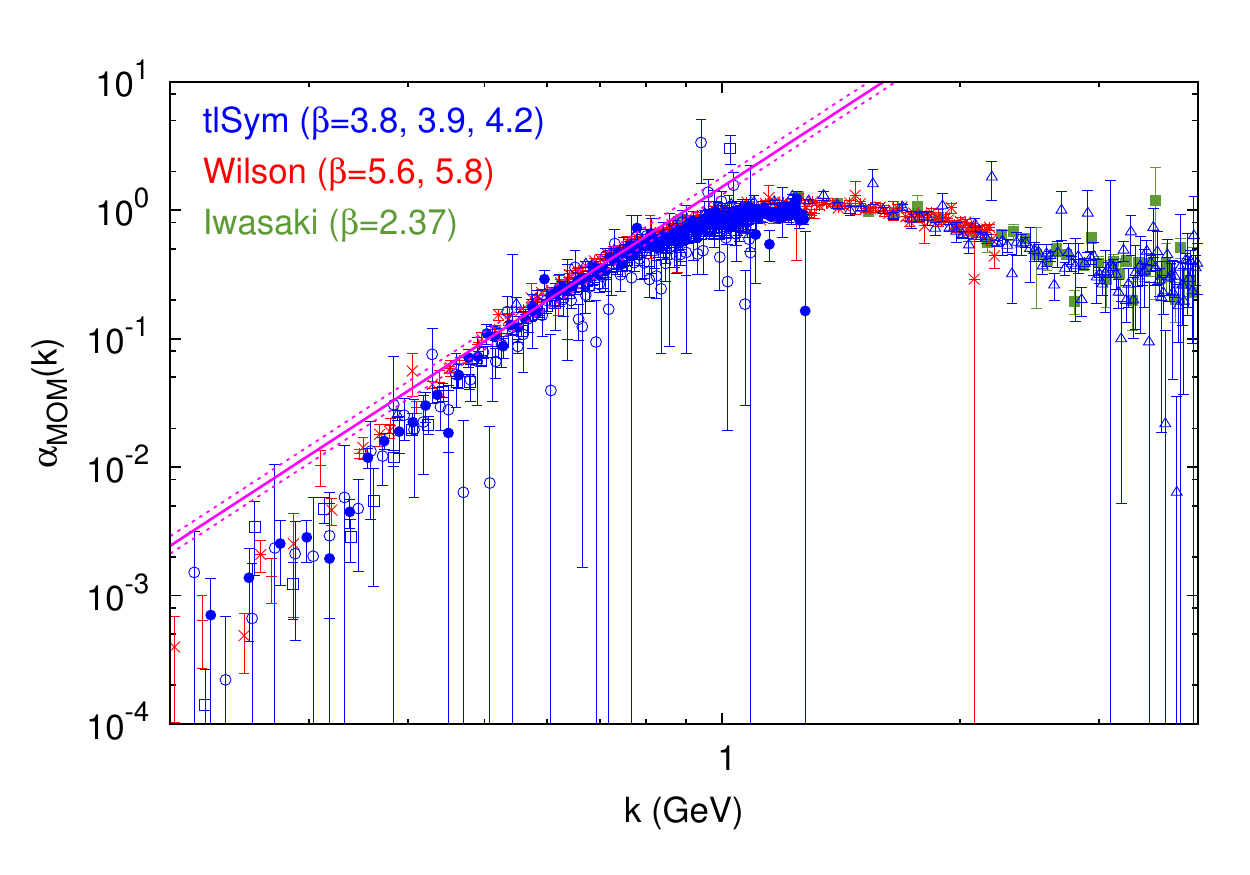} &
\includegraphics[width=0.32\textwidth]{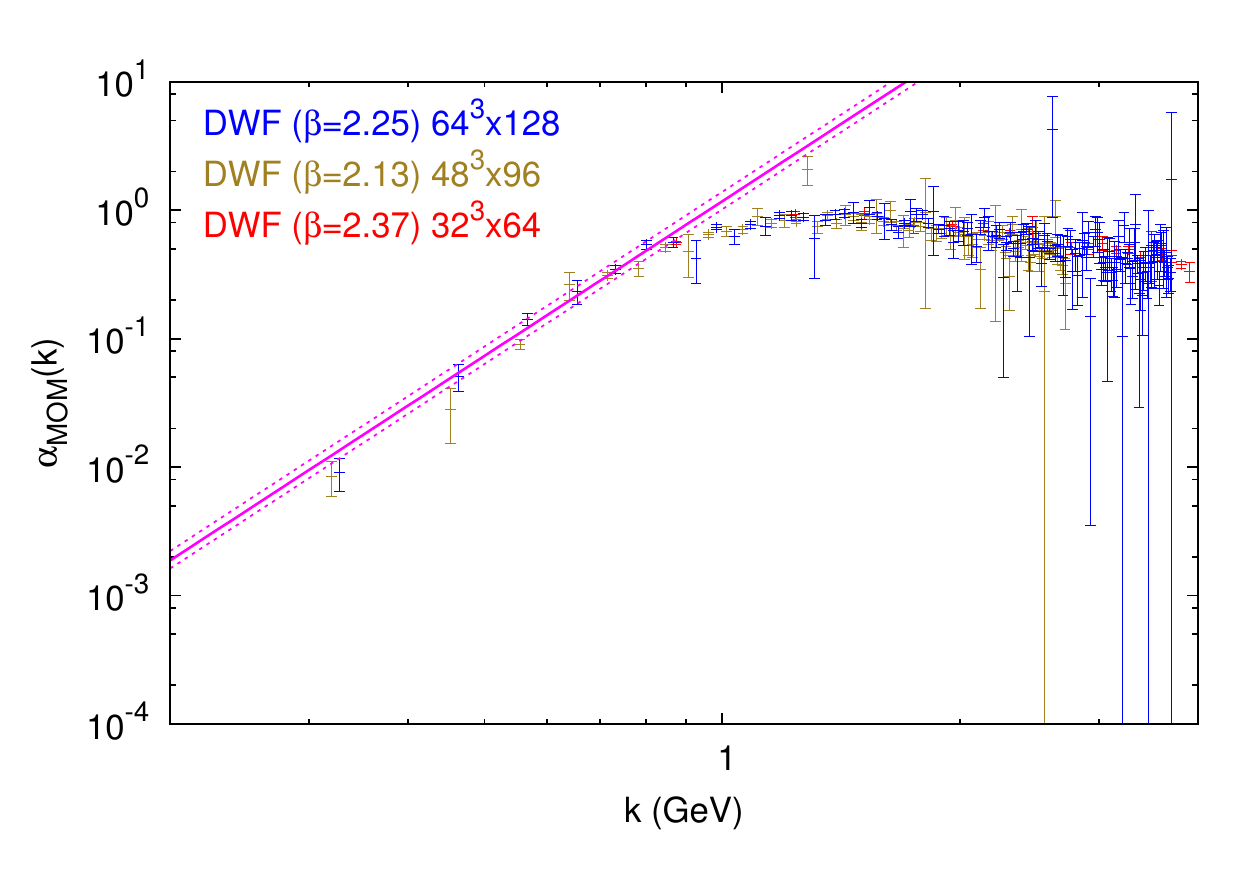} &
\includegraphics[width=0.32\textwidth]{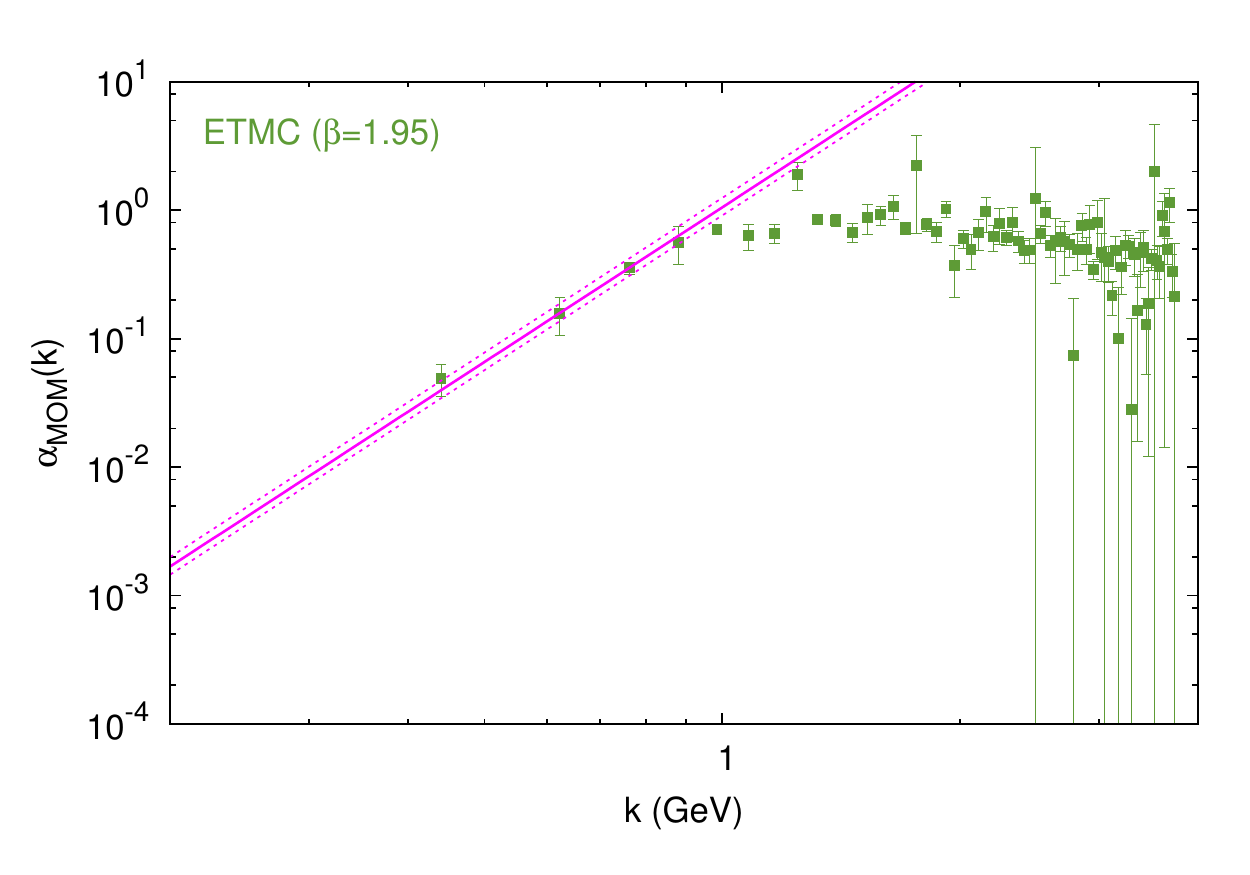}
\end{tabular}
\end{center}
\caption{(Color online) $\alpha_{\rm MOM}$ vs the momenta obtained from quenched gauge configurations (left), $N_f=2+1$ Domain-Wall (center) and ETMC $N_f=2+1+1$ (right). The fit to the instanton prediction is represented by the straight line. }
\label{fig:alpha}
\end{figure}

Without the application of any filtering technique, the running at large momenta is dominated by the perturbative prescription, giving rise to the appearance of asymptotic freedom, but the deep non-perturbative region (typically below $\sim 1{\rm GeV}$) exhibits a behavior that fits to Eq.(\ref{eq:k4}). This finding serves as a confirmation that the low momenta (large distance) correlations are dominated by instanton-like objects, and allows to extract the instanton density without need for any filtering technique. The running of $\alpha_{\rm MOM}(k)$ obtained from the lattice data has been plotted in Fig.~\ref{fig:alpha} for $N_f=0$ (left), $N_f=2+1$ (middle) and $N_f=2+1+1$ (right). The running between $0.3$ and $0.9$ GeV fits in all of them to the power-law given by Eq.~\ref{eq:k4} (straight line in the plots). Indeed, the nice scaling of this quantity in the deep IR regime for different $\beta$'s make it useful for a determination of the lattice spacing, at least in the case of pure Yang-Mills theories. When the lattice volume is large enough and momenta below $0.3$ GeV are available, this instanton prediction breaks down presumably due to the neighborhood of a zero crossing for the three gluon vertex ~\cite{ATHENODOROU2016444,PhysRevD.95.114503}. From the point of view of the ILM, the failure for small momenta is expected because for large distances (typically much larger than the instanton size) the model of uncorrelated instantons is not expected to work; in any case the purely quantum nature of the zero-crossing discussed in~\cite{ATHENODOROU2016444,PhysRevD.95.114503} seems to be in contradiction with an instantonic explanation. More work should be devoted to this issue in the future.

%\begin{figure}[h!]
%\begin{center}
%\begin{tabular}{c}
%\includegraphics[width=0.55\textwidth]{alpha1.pdf} \\
%\includegraphics[width=0.55\textwidth]{alpha_DWF.pdf} \\
%\includegraphics[width=0.55\textwidth]{alpha_ETMC.pdf} 
%\end{tabular}
%\end{center}
%\caption{(Color online) $\alpha_{\rm MOM}$ vs the momenta obtained from quenched gauge configurations (top), $N_f=2+1$ Domain-Wall (center) and ETMC $N_f=2+1+1$ (bottom). The fit to the instanton prediction is represented by the straight line.}
%\label{fig:alpha}
%\end{figure}

The fit of the deep IR running of $\alpha_{\rm MOM}(k)$ 
suggests that the instanton density increases monotonously with the number of dynamical flavors. If we 
furthermore accept that the ratio $\delta\rho^2/\bar\rho^2$ is the same for $N_f=0$, $N_f=2+1$ and $N_f=2+1+1$, we obtain that the instanton density increases by a factor $1.3(1)$ from $N_f=0$ to $N_f=2+1$ and by a factor $1.5(1)$ from $N_f=0$ to $N_f=2+1+1$. The fact that instanton density grows with the presence of dynamical flavors is already known~\cite{PhysRevD.78.054506}, although the role of the charm quark and the dependence on the quark masses has not been studied in detail yet. The density for the quenched case can be obtained once the factor $c$ in Eq.~(\ref{eq:k4}) has been fixed. Accepting a ratio $\delta\rho^2/\bar\rho^2\approx 0.015$ taken from \cite{PRD58Teper}, the density for $N_f=0$ would be $n\approx 12(2) {\rm fm}^{-4}$. 

\section{Direct identification of instantons from topological charge density}
\label{sec:localization}

\subsection{Topological charge density}

From any gauge field configuration, the field strength tensor, $F^a_{\mu\nu}(x)$ can be computed from the plaquette $\square_{\mu\nu}(x)$ as:
\begin{equation}
-i a^2 g F^a_{\mu\nu}(x) = {\rm Tr}\left[\left(\frac{\square_{\mu\nu}(x)-\square^\dagger_{\mu\nu}(x)}{2} - \frac{1}{3}{\rm Tr}\frac{\square_{\mu\nu}(x)-\square^\dagger_{\mu\nu}(x)}{2}\right)\lambda^a \right]\ .
\end{equation}
The normalized\footnote{Each instanton has an action of $8\pi^2$, which we included in the normalization so that $\int S(x) d^4x = 1$ and $\int Q(x) d^4x = \pm 1$.} action density and topological charge density can then be obtained as $S(x)=\frac{1}{32\pi^2}F_{\mu\nu}^a F_{\mu\nu}^a$ and $Q(x)=\frac{1}{32\pi^2}F_{\mu\nu}^a \widetilde{F}_{\mu\nu}^a$, with $\widetilde{F}_{\mu\nu}^a=\frac{1}{2}\epsilon_{\mu\nu\rho\sigma}{F}_{\rho\sigma}^a$. We used the so-called clover prescription for the calculation of the action and topological charge densities from the lattice gauge fields that computes the averages of the four plaquettes belonging to each site for each plane.

Near an instanton (anti-instanton), and assuming a BPST profile, the action and topological charge densities should behave as:
\beq
S(x) = \pm Q(x)= \frac{6}{\pi^2\rho^4} \left(\frac{\rho^2}{(x-x_0)^2+\rho^2}\right)^4
\label{eq:bpst}
\eeq
with ``$+$'' sign for instantons and ``$-$'' sign for anti-instantons. Nevertheless, the calculation of $S(x)$ or $Q(x)$ 
shows a strong dominance of short range, or UV, fluctuations and only after the application of some filtering technique for eliminating those UV fluctuations smooth lumps similar to \eq{eq:bpst} are revealed. In the following subsection, we will briefly describe the application of Wilson-Flow as a filter for eliminating UV fluctuations.

\subsection{Wilson flow}
\label{sec:WF}

While Wilson flow (WF) shares the capacity of eliminating the short-distance fluctuations with many other techniques such as cooling or smearing, it has solid theoretical foundations analyzed in
%a theoretical ground with attractive features such as the simple renormalization of the flown fields
\cite{Luscher2011}. We refer the reader to the original literature for all the details of the WF.

% In the continuum language, the field after a flow time $\tau$, $B_\mu(\tau,x)$ is obtained as the solution of the following first order differential equation:
%\beq
%\frac{\partial B_\mu}{\partial \tau} = D_\mu G_{\mu\nu}\ ,
%\eeq
%where $D_\mu$ and $G_{\mu\nu}$ are the covariant derivative and field strength tensor for the flown fields respectively. As initial condition for the differential equation, $B_\mu=A_\mu$ is set, so that for $\tau=0$ we start with the original gauge fields that are modified as the flow time grows. The formal solution of the previous equation is:
%\beq
%B_\mu(\tau,x)=\int d^4y \frac{e^{-(x-y)^2/4\tau}}{(4\pi\tau)^2} A_\mu(x)
%\eeq
%whose qualitative effect is that of suppressing short-distance fluctuations up to a distance of $\sqrt{8\tau}$.

Despite its important features of well defined continuum limit, smoothness of the solutions, etc., the effect of WF 
over the gauge fields is equivalent to that of cooling, at least for the determination of the topological charge~\cite{PhysRevD.89.105005,PhysRevD.92.125014}.
A very interesting property of the WF is that an isolated instanton (or more properly any exact solution of $D_\mu G_{\mu\nu}=0$) is a steady state solution of the flow equation. In this sense, WF may be thought as a filtering technique that preserves instantons. In practice this is not completely true for two different reasons: first because the whole argument corresponds to the continuum formulation of both instantons and flow, and second because isolated instantons are not an adequate description of the QCD vacuum, where a rather dense distribution of them is expected~\cite{PhysRevD.88.034501}.

Following \cite{Luscher2010}, we will write flow time in units of $t_0$, defined by $\sqrt{8t_0}=0.3 {\rm fm}$. This allows to settle physical units for the Wilson flow times for different lattices, given that $t_0=a^2\tau_0=0.01125 {\rm fm}^2$. Thus, the flow time in physical units will be fixed as $t=\tau t_0 /\tau_0$.

\subsection{Locating instantons}
\label{sec:locating}

\begin{figure}[h!]
\begin{center}
\includegraphics[width=0.55\textwidth]{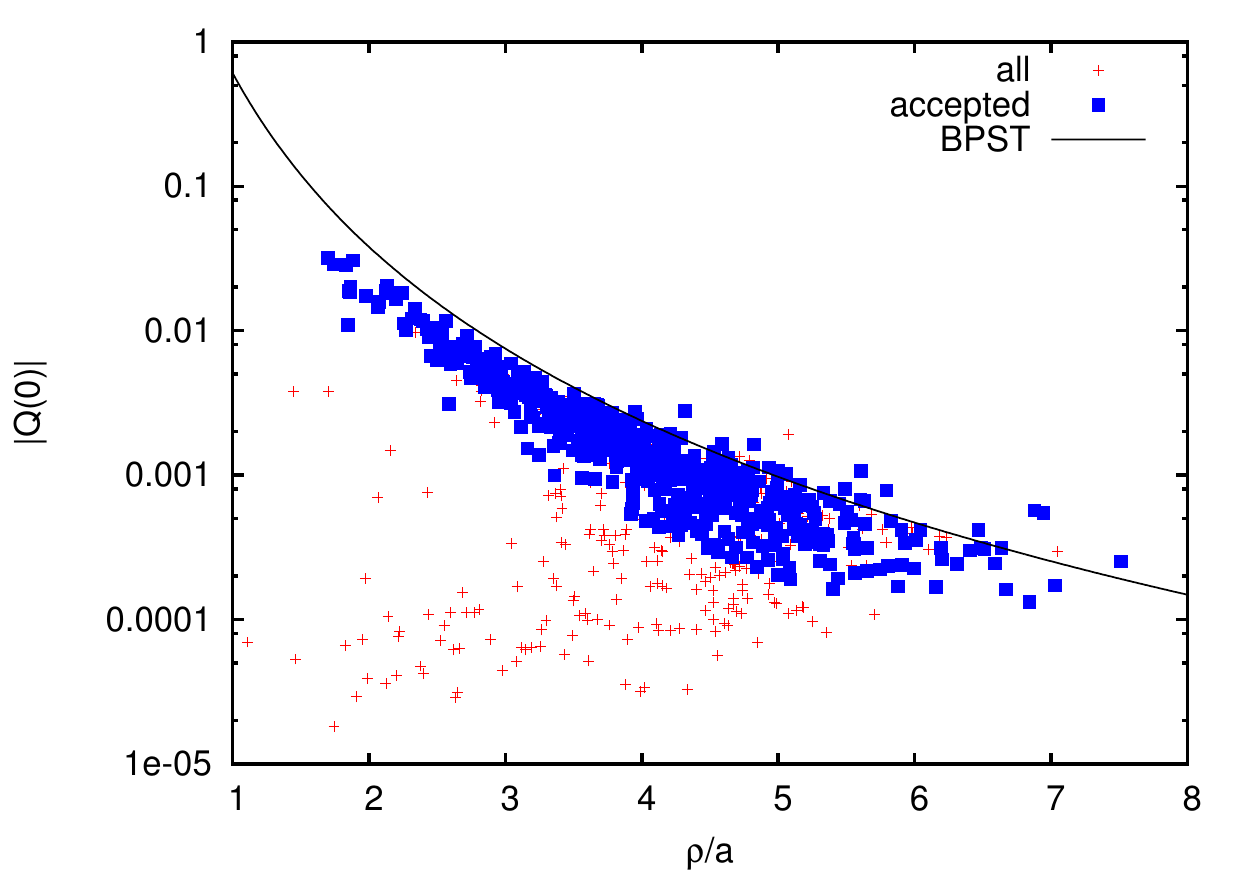} 
\end{center}
\caption{(Color online) Absolute value of the topological charge density at the extrema vs instanton size $\rho$ determined from the fit described in the text for a particular gauge field configuration of $\beta=4.2$, $V=32^4$ after $t/t_0=13.7$ Wilson flow. Those extrema accepted as instantons are represented by blue squares, while the rejected ones are represented by red crosses. The full line represents the relation \eq{eq:bpst}.}
\label{fig:q_vs_rho}
\end{figure}

After the application of WF, the short distance fluctuations present in both the action and topological charge densities are suppressed and smooth lumps are unveiled. To distinguish remaining short range fluctuations from actual instantons, we will start by finding the local extrema of the topological charge density, defined as those sites $x_0$ where $Q(x_0)$ is larger (smaller) than the closest 8 neighbors. Due to the persistence of some fluctuations after WF, not all the extrema of the topological charge density correspond to instantons. For discriminating instantons, we considered the ratio of the topological charge at the closest 8 neighbors to the value at the center that, according to  Eq.(\ref{eq:bpst}) should be
\beq
\frac{Q(x)}{Q(x_0)} = \left(\frac{\rho^2}{(x-x_0)^2+\rho^2}\right)^4
\eeq
and fitted the lattice data to this expression using $\rho$ as a free parameter. For a BPST instanton, the value of the topological charge at the instanton center is related to the instanton size by $Q(x_0)=\frac{6}{\pi^2\rho^4}$. To check this dependency, the peak value of the topological charge density has been plotted versus the fitted size in Fig.~\ref{fig:q_vs_rho} for a typical quenched configuration after $t/t_0=13.7$ Wilson flow. In this plot there is a point for each local extremum of the topological charge density, while the continuum line corresponds to Eq.~(\ref{eq:bpst}). Most of the extrema are disposed near the continuum line, although a non-negligible number of outliers have to be discarded. For larger flow times, the extrema tend to concentrate close to the continuum line. Furthermore, when an extremum  at site $x_i$ is identified as an instanton of size $\rho_i$, no other candidate is accepted at distances smaller than $\rho_i$.
That means that in the core of an instanton we assume there is no possibility of finding another instanton or, in other words, that the instanton core is not distorted by the presence of other instantons (other methods in the literature for localizing instanton-like objects like \cite{PhysRevD.78.054506} fix a minimum separation between instantons).

\begin{figure}[h!]
\begin{center}
\begin{tabular}{c c}
\includegraphics[width=0.45\textwidth]{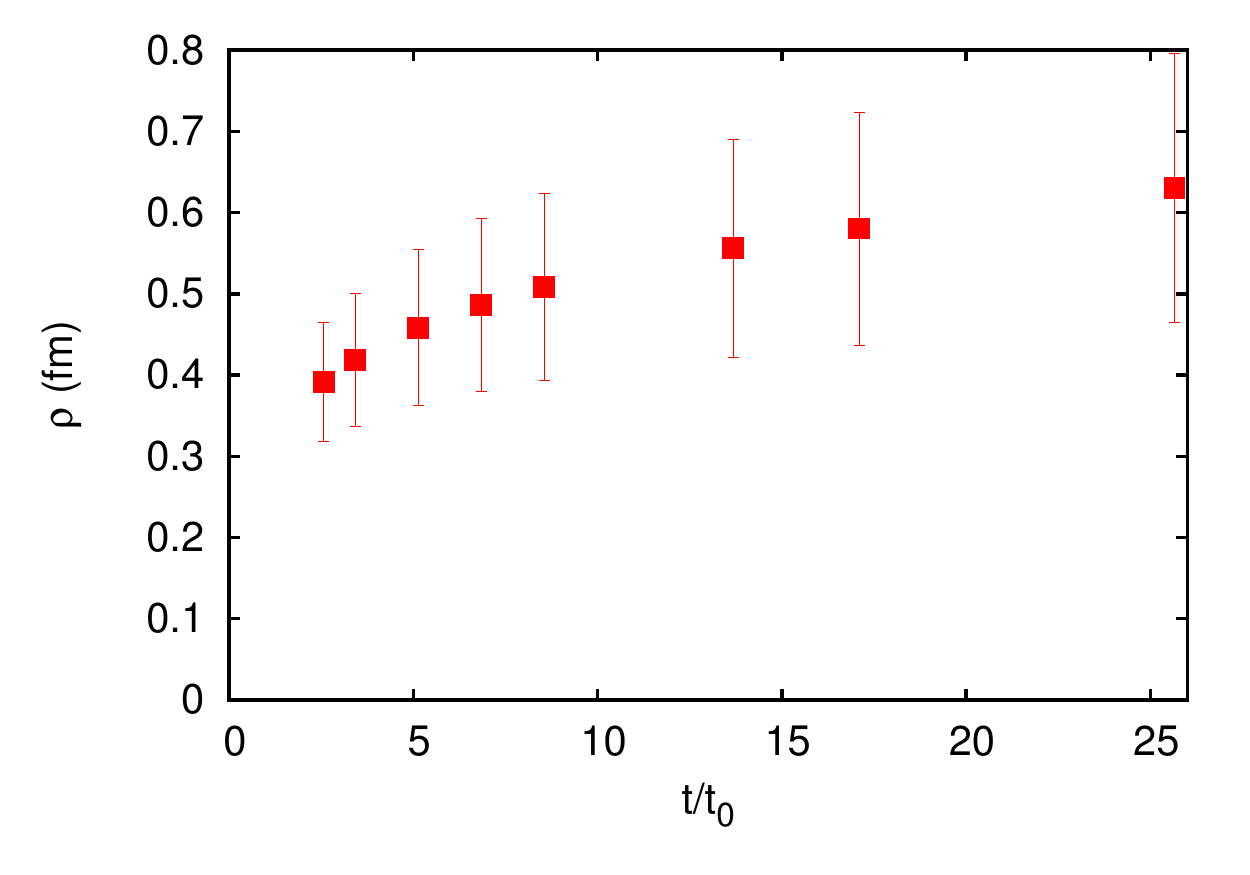} &
\includegraphics[width=0.45\textwidth]{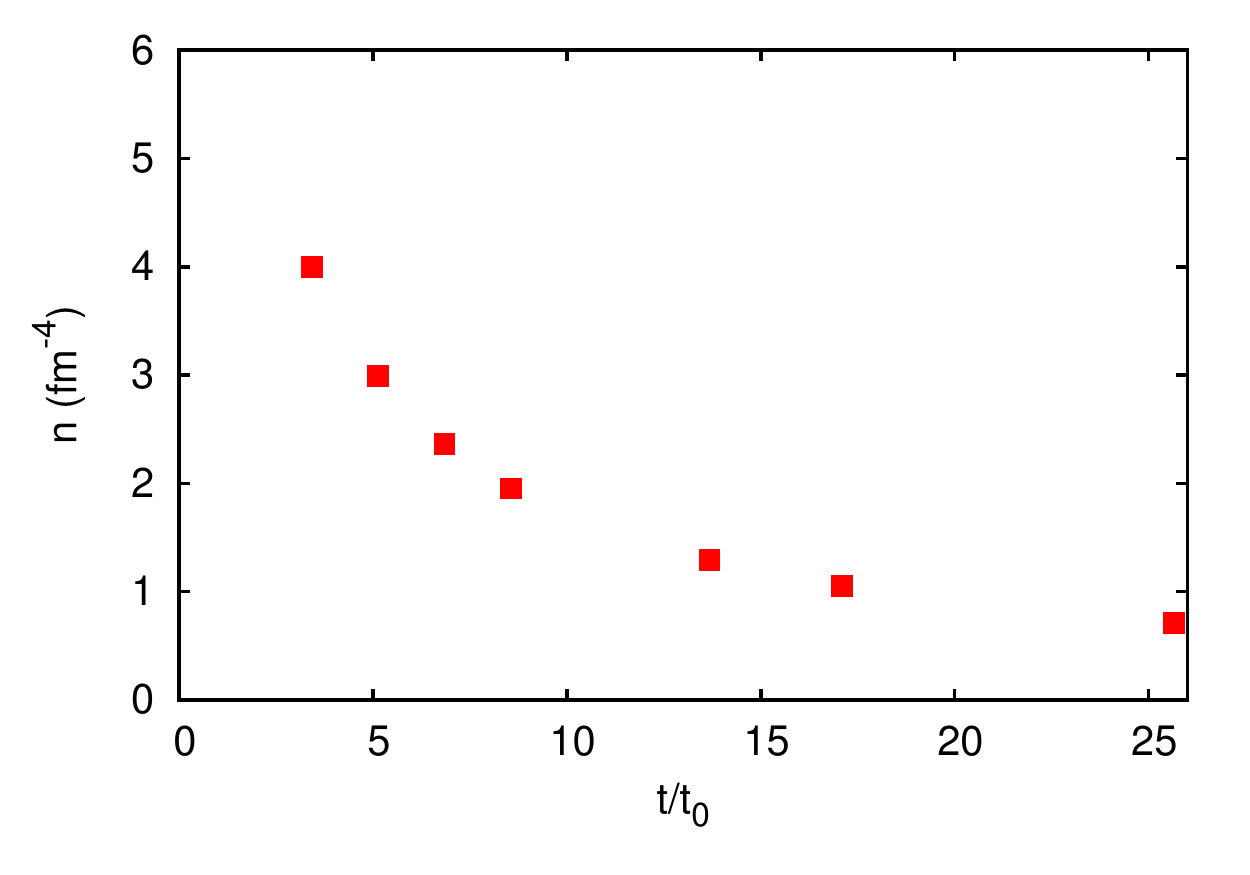}
\end{tabular}
\end{center}
\caption{(Color online) Evolution of the average instanton size (left) and density (right) with Wilson flow time for 100 quenched gauge field configurations corresponding to $\beta=4.2$. In the left plot, the error bars show the width of the instanton size distribution rather than the standard deviation of the mean.}
\label{fig:size_density_Wilson}
\end{figure}

Although the fact that the parameters of the instanton ensemble evolve with the suppression of UV fluctuations is known~\cite{GARCIAPEREZ1994535,NEGELE199992,HART2001280,PhysRevD.88.034501,PhysRevD.89.105005}, %many references to be added
few authors have studied their evolution in detail, quoting just the values of instanton density, or size distribution after a fixed flow time, cooling or smearing. The picture that emerges from Fig.~\ref{fig:size_density_Wilson} is that of a very dense ensemble of instantons with small size that is more diluted as Wilson flow eliminates instantons and anti-instantons while the remaining ones become larger with the flow.

The instanton densities we have measured are $n \sim 1 {\rm fm}^{-4}$ for rather large flow times, a value that agrees with the one that can be set from the gluon condensate $\VG$, and that has served as reference for decades \cite{RevModPhys.70.323}. Nevertheless, if we try to infer the density at zero Wilson flow time from the results in Fig.\ref{fig:size_density_Wilson}, the density results much larger. Although it is difficult to extrapolate back to $\tau=0$ from our results, it points towards an order of magnitude larger.

Concerning the instanton size, Garcia-Perez et al. \cite{GARCIAPEREZ1994535} found that with Wilson action, individual instantons should shrink under a cooling procedure while with overimproved actions they stabilize or grow with the application of cooling. 
Although it is not fully known how the use of different filtering techniques may modify their conclusions, the fact of being in a dense liquid instead of isolated is for sure a dramatic change respect to their conditions. Indeed we found that most of the instantons we localized grow with the application of the Wilson flow, while only small instantons shrink (and eventually disappear).
We checked that if the gradient flow is implemented via the Iwasaki gauge action, the observed evolution does not differ from the one described above.

%
%\begin{figure}[h!]
%\begin{center}
%\includegraphics[width=0.55\textwidth]{all_hist.pdf}
%\end{center}
%\caption{(Color online) Histogram of the instanton sizes (normalized to the instanton density) for different Wilson flow %times for 100 quenched gauge field configurations at $\beta=4.2$.}%
%\label{fig:all_hist}
%\end{figure}

%\begin{figure}[h!]
%\begin{center}
%\begin{tabular}{c c}
%\includegraphics[width=0.45\textwidth]{{gr_tau=2.0}.pdf} &
%\includegraphics[width=0.45\textwidth]{{gr_tau=4.0}.pdf} \\
%\includegraphics[width=0.45\textwidth]{{gr_tau=8.0}.pdf} &
%\includegraphics[width=0.45\textwidth]{{gr_tau=15.0}.pdf}
%\end{tabular}
%\end{center}
%\caption{(Color online) Pair correlation function for $II$ and $AA$ pairs (red) and for $IA$ pairs (blue) for $\beta=4.2$, $V=32^4$ and a Wilson flow time of $t/t_0=3.4$, $6.9$, $13.7$ and $25.6$. The distribution is characteristic of a gas, with an excluded volume that is larger for $IA$ pairs than for $II$ or $AA$. The appearance of a peak for $IA$ pairs for small flow times (top row) shows that each instanton tends to be surrounded by a shell of opposite topological charge.}
%\label{fig:gr}
%\end{figure}

\subsection{Running of $\alpha_{\rm MOM}(k)$ after Wilson flow}

After removing short-distance fluctuations, the ratio in \eq{eq:mom} can be computed again (with the additional cost of Landau gauge fixing at each flow time). Indeed, it has been found that the $k^4$ prediction of the ILM reproduces the lattice data for $\alpha_{\rm MOM}(k)$ both in the large and small momenta regions. This has been tested using different methods for filtering  UV fluctuations such as cooling~\cite{JHEP04Boucaud} and Wilson flow~\cite{ATHENODOROU2016354}. The slope of $\alpha_{\rm MOM}(k)$ in $k^4$ allows a direct computation of the instanton density from the lattice data at large momenta, that can be compared to the densities obtained from a direct instanton counting using the localization algorithm described above.

The densities obtained for an ensemble of $100$ quenched gauge field configurations at $\beta=4.2$  from the analysis of the running and the localization algorithm have been included in table \ref{tab:comparison} for three different flow times. A comparison of the densities obtained from the two methods shows that the localization systematically finds less instantons than the expected from their influence on the running. This may be due to some instantons that are not recognized by the algorithm, possibly due to the fact that close instantons are difficult to detect. The possibility of a systematic overestimation of the densities extracted from the running cannot be excluded neither, and the reason for this $\sim 25\%$ discrepancy between the  densities measured with both methods requires further scrutiny.

\begin{table}[h!]
  \small
  \centering
  \caption{Comparison between the instanton densities obtained from the running of $\alpha_{\rm MOM}(k)$ for large momenta (data taken from \cite{ATHENODOROU2016354}) and the direct localization as a function of the flow time. The errors quoted for the localization algorithm are just the standard deviation of the densities obtained from the different configurations.}
  \label{tab:comparison} % Give a unique label
  \begin{tabular}{l l l}\toprule
  $t/t_0$  & Running $\alpha_{\rm MOM}(k)$ & Localization  \\\midrule
  6.8  & 3.5(1) & 2.445(5) \\
  13.7 & 1.75(4) & 1.351(4) \\
  25.6 & 0.98(5) & 0.744(3) \\\bottomrule
  \end{tabular}
\end{table}

The transition between the two regimes of small and large momenta given by \eq{eq:k4} occurs at a scale driven by $\rho^{-1}$. The observation of a shift towards smaller values of the momenta with WF made in \cite{ATHENODOROU2016354}, is therefore an indirect evidence of instanton size enlarging, as the one stated here thanks to the direct localization algorithm presented.

\section{Conclusions}

The IR running of $\alpha_{\rm MOM}(k)$ for $N_f=0$,  $N_f=2+1$, and $N_f=2+1+1$ has been analyzed, and fits nicely to an uncorrelated instanton liquid model for momenta $k\in (0.3,0.9)$ GeV. From the fit, we found that the instanton density in the dynamical configurations is higher than in the quenched ones, increasing with the number of dynamical quark flavors. 
The fact that isolated instantons are associated to exact zero modes of Dirac operator make less likely to find instantons {\it isolated} in the presence of dynamical quarks because zero modes of the Dirac operator would have zero weight (at least in the chiral limit), and would be avoided by the Monte Carlo simulations. 
The observed growth of the density would serve to limit the possibility of finding instantons isolated. The value of the instanton density inferred from the fits would be rather high, being above $10 {\rm fm}^{-4}$ already in the quenched case and higher for  $N_f=2+1$ and $N_f=2+1+1$ (albeit the comparison may be biased by a slight systematic deviation of the physical scale setting as the one discussed in \cite{PhysRevD.96.098501}).

We presented an algorithm for localizing instanton patterns from the topological charge distribution, and quantified the instanton content of each gauge field configuration for a set of $\beta=4.2$ quenched configurations after the application of Wilson flow, measuring instanton sizes and densities. We observed that as the flow time grows, the instanton ensemble gets more diluted while the instanton size grows monotonously. A combination of both approaches for obtaining instanton densities after Wilson flow shows that, despite the differences between both methods, there is a strong dependence of the instanton density on the filtering done (flow time).

\section{Acknowledgements}
 {\small
We thank the ETM and RBC/UKQCD collaborations for sharing their configurations with us.
SZ acknowledges
support by the National Science Foundation (USA) under
grant PHY-1516509 and by the Jefferson Science Associates,
LLC under U.S. DOE Contract \# DE-AC05-
06OR23177.}

%\clearpage
\bibliography{lattice2017}

%%%%%%%%%%%%%%%%%%%%%%%%%%%%%%%%%%%%%%%%%%%%%%%%%%%%%%%%%%%%%%%%%%%%%%%%%%%%%
\end{document}